\documentclass[12pt]{iopart}
\usepackage{iopams}  
\usepackage{amstext}

\usepackage{graphicx,color}

\begin{document}

\title[Geometric distortion of area in medical ultrasound images]{Geometric distortion of area in medical ultrasound images}

\author{T Bland$^{1}$, J Tong$^2$, B Ward$^2$ and N G Parker$^1$}

\address{$^1$School of Mathematics and Statistics, Newcastle University, Newcastle upon Tyne, UK}

\address{$^2$Regional Medical Physics Department, Freeman Hospital, Newcastle upon Tyne, UK}
\ead{nick.parker@newcastle.ac.uk}
\vspace{10pt}

\begin{abstract}
Medical ultrasound scanners are typically calibrated to the soft tissue average of 1540 m s$^{-1}$.  In regions of different sound speed, for example organs and tumours, the B-mode image then becomes a distortion of the true tissue cross-section, due to the misrepresentation of length and refraction.  We develop a ray model to predict the geometrical distortion of a general elliptical object with atypical speed of sound, and quantify the ensuing image distortion as a function of the key parameters, including the speed of sound mismatch, the object size and its elongation.  Our findings show that the distortion of area can be significant, even for relatively small speed of sound mismatches.  Our findings are verified by ultrasound imaging of a test object. 
\end{abstract}

\vspace{2pc}

\section{Introduction}
\label{intro}

Medical ultrasound imaging is routinely used to evaluate length, area and volume as part of clinical diagnostic assessment, for example in sizing of organs and tumours, and assessing the age, weight and normality of fetuses \cite{Rumack}.  The images are obtained through ultrasound time-of-flight measurements, converted into the spatial domain via an assumed speed of sound, typically taken to be the soft tissue average of $1540$ m~s$^{-1}$.  However, bodily soft tissue is acoustically inhomogeneous and its speed of sound can vary by several percent across different regions \cite{Fontanarosa}.  For example, typical speeds of sound in fat, muscle and liver are $1440$, $1580$ and $1590$ m s$^{-1}$, respectively \cite{Duck}.   The image is then geometrically distorted from the actual tissue profile, where the variable speed of sound is not accounted for.  This, in turn, leads to systematic inaccuracies in the evaluation of length, area and volume of anatomical structures.

To illustrate this effect, consider an ambient medium with speed of sound $v_1$ and an ultrasound scanner which assumes the same value.   An object with atypical speed of sound $v_2 \neq v_1$ is embedded within the medium and a B-mode image taken of its cross-section.  Then, the distance within the object will be misrepresented in the image and the refraction at the object boundary will not be accounted for, both of which will contribute to the geometric distortion of the object's image.  While this effect has been predicted previously \cite{Sommer,Robinson,Ziskin1,Ziskin2,Rubin,Steel2004}, a quantitative understanding is lacking.  Such understanding is important in establishing errors in clinical measurements, and devising corrective strategies.  Take, for example, the evaluation of the volume of organs and tumours now possible through three-dimensional ultrasound imaging \cite{Campani_1998,Gilja_1999,Treece_2001}.  This capability has diagnostic importance, particularly in charting the evolution of a disease, and has been demonstrated on a variety of structures including the bladder \cite{Riccabona1,Riccabona2,Kristiansen,Suwanrath_2009}, prostate \cite{Tong_2006}, kidney \cite{Yoshizaki_2013}, thyroid \cite{Lyshchik_2004} and tumours \cite{Hornblower_2007}.  The 3D image is typically reconstructed from 2D images \cite{Linney_1999,Fenster_2001}; then, if the object in question has atypical speed of sound, the above 2D distortion will introduce errors in the calculated area of each scanning plane, and thus the total calculated volume.  To date, this distortion remains largely unexplored as a source of error in area and volume measurements.

In this work we make a quantitative study of the geometric distortion arising in 2D ultrasound images of a general elliptical object with atypical speed of sound.  We develop a ray model to predict this distortion (presented in Section \ref{sec:theory}), applicable across the four main multielement transducers currently in medical use (linear, curved, phased and vector arrays).  Our aim is not to predict the full intensity distribution of the ultrasound images; to do so would require a sophisticated model of the acoustic field to take into account wave effects (such as diffraction and interference) and the nature of the beam line (such as its finite width and focussing). Our aim instead is to predict and understand the underlying mapping of the object boundary to the image boundary; as such, the ray approach is well-suited to this aim.  Its simplicity also lends to future development of strategies to correct the distortion.  Comparison to ultrasound images of phantoms (described in Section \ref{sec:exp}) confirm the success of the ray model in predicting the shape of the image.  We analyse and quantify the distortion as a function of the speed of sound ratio $v_2/v_1$, the object's shape and size, and the array type.  Our results show that a small speed of sound mismatch can lead to significant distortions: for example, a  5\% mismatch in the speeds of sound can lead to around 15\% error in the area of a circular object, and this is sensitive to the size and shape of the object.

\section{Theory}
\label{sec:theory}

\subsection{Ray Model} 
\label{sec:raz_ellipt}

We consider an elliptical (two-dimensional) object in the $x$-$z$ plane, with uniform speed of sound $v_2$, embedded in an ambient medium with uniform speed of sound $v_1$.  The object has semi-widths $b$ and $c$.  The set-up is depicted in figure 3(a).   The scanner is assumed to be correctly calibrated to the ambient speed of sound $v_1$.  We will develop the model for the case of a curved transducer array; later, we will generalize to the other main array types.  
In such an array, beam lines can be considered to spread out from a virtual source within the (curved) transducer, emanating normal to the transducer surface \cite{Goldstein2000}. The origin is taken to be the virtual source.  For generality we allow for a lateral offset $a$ between the transducer and the ellipse.  The front face of the transducer lies a distance $d$ from the source.  The object boundary is then represented by the elliptical line $z_{\rm ob}(x)$ which satisfies the relation
\begin{eqnarray}
\frac{(x-a)^2}{b^2}+\frac{\left(z_{\rm ob}-c-d-d_0\right)^2}{c^2}=1.
\label{eq:yellipse}
\end{eqnarray}
\begin{figure}[t]
\centering
\includegraphics[width=0.45\columnwidth]{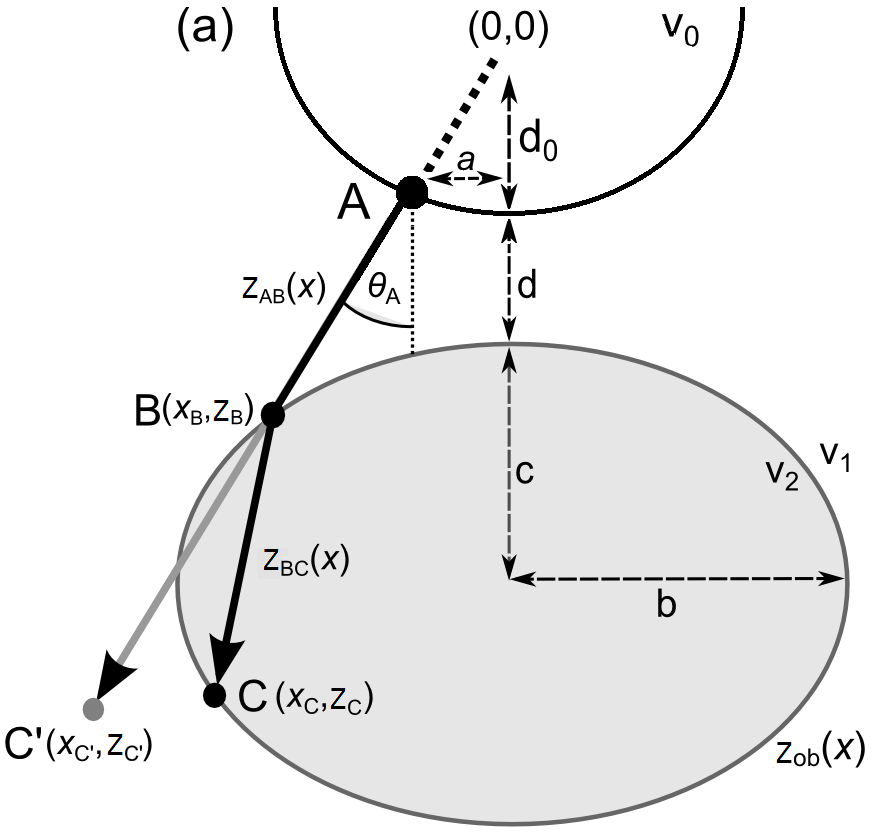}\hspace{1cm}
\includegraphics[width=0.35\columnwidth]{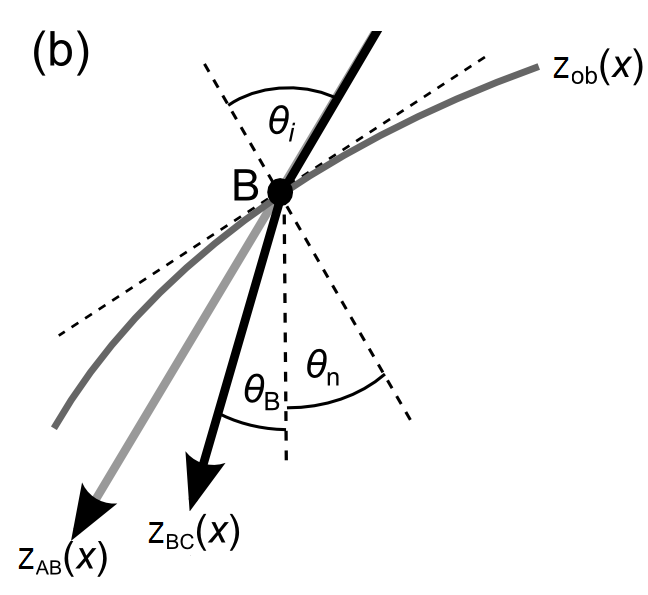}
\caption{(a) Schematic of the curved transducer array and elliptical object (speed of sound $v_2$) embedded within an ambient medium (speed of sound $v_1$).   For $v_2 \neq v_1$, refraction and misrepresentation of length within the object cause a point $C$ on the far boundary to appear at $C'$. (b) A close-up of the real ray (black) and image ray (grey), and the angles involved, at the point $B$.}
\label{fig:ellipse}
\end{figure}
We model the distortion of the object by treating the beam lines as rays.  This approximation is valid in the far-field of the source and for objects much larger than the acoustic wavelength.  Also, where the beam is incident on a boundary, the ray approach ignores mode conversion to interface waves that might occur at certain crucial angles.  Nonetheless, through comparison to experimental images, we will see that the ray model accounts for the geometric distortion in the image.

A ray is emitted from the transducer surface at point $A$ with coordinates $(x_{\rm A},z_{\rm A})$ and is incident on the front face of the object at point $B$ with coordinates $(x_{\rm B},z_{\rm B})$.  This follows the straight line path $z_{\rm AB}(x)(x)=x\cot\theta_{\rm A}$, where $\theta_{\rm A}$ is the angle of the emitted ray relative to the vertical
\begin{eqnarray}
\theta_{\rm A}=\arctan\frac{x_{\rm B}}{z_{\rm B}}.
\label{eq:circlet1}
\end{eqnarray}
At $B$, the angle between the normal of the boundary and the vertical, $\theta_{\rm n}$, can calculated from the gradient of the boundary at $B$ via
\begin{eqnarray}
\tan \theta_{\rm n}=\frac{{\rm d}z_{\rm ob}}{{\rm d}x}\Big\vert_{(x_{\rm B},z_{\rm B})} = \frac{c^2 (x_{\rm B}-a) }{b^2(d+d_0+c-z_{\rm B})}.
\label{eq:circletn0}
\end{eqnarray}

The angle of incidence at $B$ (relative to the normal) is then $\theta_{\rm i}=\theta_{\rm n}+\theta_{\rm A}$.

At the front face, a component of the wave will be back-scattered to the transducer, registering the presence of this point in the image.  Since this ray travels only through the ambient medium, for which the scanner is assumed to be calibrated, the image of this point corresponds to its real location.  Of more interest to us is the component of the wave which is refracted through the boundary (the components of the wave that undergo specular reflection or scattering away from the transducer are not registered by the transducer and so we ignore them).  This ray is incident on the object's back face at point $C$, with coordinates $(x_{\rm C},z_{\rm C})$, and represented by the straight line $z_{\rm BC}(x)$,
\begin{eqnarray}
z_{\rm BC}(x)=(x-x_{\rm B})\cot \theta_{\rm B} +z_{\rm B}.
\label{eq:yray}
\end{eqnarray}
$\theta_{\rm B}$ is the angle from the vertical direction, given by $\theta_{\rm B}=\theta_{\rm r}-\theta_{\rm n}$, where $\theta_{\rm r}$ is the angle of refraction.  To obtain an expression for $\theta_{\rm r}$, we rearrange Snell's law, $\displaystyle \sin \theta_{\rm i} / \sin \theta_{\rm r}=v_1/v_2$, to give
\begin{eqnarray}
\theta_{\rm r}=\arcsin\left[\frac{v_2}{v_1}\sin(\theta_{\rm A}+\theta_{\rm n})\right].
\label{eq:circlet2}
\end{eqnarray}
To establish $(x_{\rm C},z_{\rm C})$, we seek the intersections of the refracted ray $z_{\rm BC}(x)$ with the boundary $z_{\rm ob}(x)$.  Rearranging equation (\ref{eq:yray}) for $x$ and inserting into equation (\ref{eq:yellipse}) gives the relation
\begin{equation}
\frac{\left[(z-z_{\rm B})\tan \theta_{\rm B}+x_{\rm B}-a\right]^2}{b^2}+\frac{\left[z-c-d-d_0\right]^2}{c^2}=1. \nonumber
\end{equation}
Multiplying this out leads to a quadratic equation in $z$ and thus two solutions for $z$.  The larger of these corresponds to point $C$ (and the smaller to point $B$).  Hence we obtain the $z$-coordinate of $C$ to be

\begin{eqnarray}
\fl z_{\rm C}=z_{\rm B}
\\
\fl+2\sqrt{\left[\frac{c^2(x_{\rm B}-a)\tan \theta_{\rm B}-b^2 d'-z_{\rm B}c^2\tan^2 \theta_{\rm B}}{c^2 \tan^2 \theta_{\rm B}+b^2}\right]^2
+\frac{c^2 b^2-b^2 d'^2-c^2\left[x_{\rm B}-a-z_{\rm B} \tan \theta_{\rm B}\right]^2}{c^2\tan^2 \theta_{\rm B}+b^2}},  \nonumber
\label{eq:elly0}
\end{eqnarray}
where $d'=d+d_0+c$. $x_{\rm C}$ then follows from equation (\ref{eq:yellipse}).  

Having established the path to $C$, we proceed to locate its image, $C'$.  The paths $AB$ and $BC$ have lengths $L_{\rm AB}=\sqrt{x_{\rm B}^2+z_{\rm B}^2}$ and $L_{\rm BC}=\sqrt{\left(x_{\rm C}-x_{\rm B}\right)^2+\left(z_{\rm C}-z_{\rm B} \right)^2}$, respectively.  Assuming that the ray returns to the transducer along the same path, the total travel time is
\begin{eqnarray}
t=\frac{2 L_{\rm AB}}{v_1}+\frac{2 L_{\rm BC}}{v_2}.
\label{eq:t1}
\end{eqnarray}
The transducer, calibrated to the ambient speed of sound $v_1$, perceives the ray as the unrefracted path, i.e. along $z_{\rm AB}(x)$, with the same travel time $t$.  Hence the path $AC'$ has length
\begin{eqnarray}
L_{\rm AC'}=\frac{v_1 t}{2}=L_{\rm AB}+\frac{v_1}{v_2}L_{\rm BC}.
\label{eq:length}
\end{eqnarray}
The position of $C'$ then follows as 
\begin{equation}
(x_{\rm C'},z_{\rm C'})=(L_{\rm AC'}\sin\theta_{\rm A},L_{\rm AC'}\cos\theta_{\rm A}).
\label{eq:circleimage}
\end{equation}
For $v_1 \neq v_2$ the image point differs from the real point, partly due to refraction of the ray at the near face of the object, and secondly due to the rescaling of length due to the speed of sound misrepresentation.  For $v_2 < v_1$, the image is extended from the transducer (due to the longer travel time within the object), while for $v_2>v_1$ the image is moved closer to the transducer (due to the shorter travel time within the object).

\subsection{Generalization to Other Array Types}
\label{sec:raz_gen}

There are 3 other main types of multielement transducer currently in use in medical imaging: the linear, phased and vector arrays.  The ray structure for these array types can be found elsewhere \cite{Goldstein2000}.  Below we summarize how the ray model can be extended to these types.

\begin{description}

\item[Vector array:]
In this array, beam lines spread out from a virtual source located within the (flat) transducer.  Since we assume that $v_1=v_0$, no virtual refraction takes place at the transducer surface, and so the only modification is in the position at which the ray emerges from the surface, accounted for in the model by the coordinates ($x_{\rm A},z_{\rm A}$).

\item[Phased array:]  In a phased array beam lines spread out from a virtual source at the surface of the transducer, accounted for by setting $d_0=0$.  

\item[Linear array:]
In a linear array, beam lines emerge from the transducer surface, i.e. $d_0=0$, normal to the transducer face, i.e. $\theta_{\rm A}=0$.  Note that the image position (\ref{eq:circleimage}) then reduces to
\begin{eqnarray}
(x_{\rm C'},z_{\rm C'})&=&(0,L_{\rm AC'}),
\label{eq:linearimage}
\end{eqnarray}
i.e. the image point appears directly above the source.  
\end{description}

\subsection{Area Evaluation}
\label{sec:area}
The image boundary will, in general, be irregularly-shaped.  To evaluate the enclosed area we exploit Green's theorem \cite{Greens}.  Green's theorem states that, for a region $R$ bound by a closed path $\Gamma$, which is continuous and with no crossing points, then,

\begin{equation}
 \int \hspace{-0.2cm} \int \limits_{R} \left(\frac{\partial v}{\partial x} - \frac{\partial u}{\partial z}\right)\, \mathrm{d}x~ \mathrm{d}z=\oint \limits_{\Gamma} \left\{u~\mathrm{d}x + v~\mathrm{d}z \right\},
\label{eq:greens}
\end{equation}
where $u(x,z)$ and $v(x,z)$ are arbitrary functions, and the path of integration along $\Gamma$ is taken to be counter-clockwise.  Taking the choice of functions $u(x,z)=-z/2$ and $v(x,z)=x/2$ then the left-hand side of equation (\ref{eq:greens}) reduces to $\displaystyle \int \hspace{-0.2cm} \int_R {\rm d}x~{\rm d}z$, which corresponds to the area of the region $R$.  It then follows that we can express the area $A$ of this region via a line integral around its boundary $\Gamma$ via

\begin{equation}
\textrm{A} = \frac 12 \oint \limits_\Gamma \left(-z\, \mathrm{d}x + x\, \mathrm{d}z\right).
\label{eq:greens2}
\end{equation}
We numerically implement this line integral around the image boundary (discretised as a series of points) to obtain the image area.  We quantify the image distortion through the {\it area distortion} $A_{\rm i}/A_{\rm r}$, where $A_{\rm i}$ and $A_{\rm r}$ are the areas of the image and object.

\section{Materials and Methods}
\label{sec:exp}
\begin{figure}[t]
\centering
\textsf{(a) ~~~~~~~~~~~~~~~~~~~~~~~~~~~~~~~~~~~~~~~~~~~~~~~~~~~~(b)}~~~~~~~~~~~~~~~~~~~~~~~~~~~~~~~~~~~~~~~~~~~~~~~~~~~~~~ \\
\includegraphics[width=0.45\columnwidth]{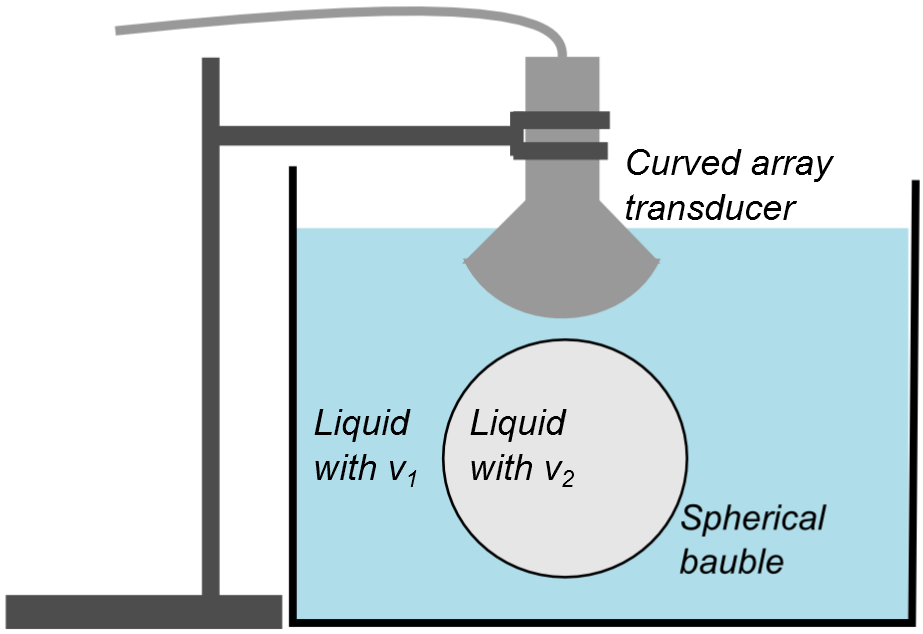} \hspace{1cm} \includegraphics[width=0.43\columnwidth]{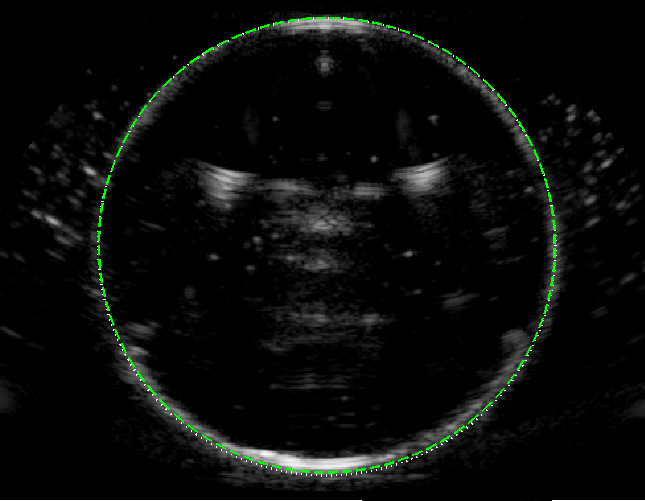}
\caption{(a) Illustration of the experimental set-up. A curved array transducer, clamped in place, images (from above) the central cross-section of a spherical bauble.  The bauble is filled with liquid of speed of sound $v_2$, and the ambient speed of sound is $v_1$.  (b) Corresponding $B$-mode image for $v_2=v_1= 1540$ m s$^{-1}$, both matched to the scanner.  The expected object boundary is shown by the dashed green line.}
\label{fig:exp}
\end{figure}

A hollow spherial aluminium bauble, filled with ``liquid 2" and sealed with a latex condom, was suspended (by a wire frame) in a bath of ``liquid 1".  Its central circular cross-section was imaged from above by an ultrasound scanner (Philips HDI 5000) and curved array (Philips C5-2 40R, with $d_0=40$ mm).  The set-up is depicted in figure~\ref{fig:exp}(a).   The bauble and condom are sufficiently thin to have negligible impact on our analysis.  The transducer is aligned manually with the equator of the object.  The overall object radius was measured with calipers as (3.99 $\pm$ 0.01) cm and the transducer-object distance $d$ determined using the scanner's electronic calipers.  The experiments were run at ambient temperature and the precise bath temperature recorded by digital thermometer.  

For the liquids we employed pure water, pure methanol and an ethanol-water mixture (with $7.25\%$ w/w ethanol).  These have room temperature speeds of sound of $1482$ m s$^{-1}$ \cite{Water}, $1100$ m s$^{-1}$ \cite{benson} and $1540$ m s$^{-1}$ \cite{Tong}, respectively (the latter being matched to the tissue average and scanner-assumed value).  The bath temperature varied in the range $18$-$23^\circ$C and we corrected the speeds of sound used in the model using the established temperature-speed relations \cite{Water,benson,Tong}.

As an initial validation we filled both the object and bath with ethanol-water mixture, such that the system has uniform speed of sound throughout, $v_2=v_1 =1540$ m s$^{-1}$.  The $B$-mode, shown in figure \ref{fig:exp}(a), shows a circular image boundary, confirming the absence of geometric distortion, and matches well to the expected circular shape of the object (green dashed line).  Note the intensity distortion around the boundary: the poles of the object are brightest as a result of specular reflection.  

Three variations of this system were considered.
\begin{description}
\item[$v_2<v_1$:]  Taking ethanol-water mixture as liquid 1 and pure water as liquid 2 embodies a system with $v_2<v_1$ ($v_2 \approx 0.96 v_1$).  

\item[$v_2\ll v_1$:]  Taking the ethanol-water mixture as liquid 1 and now methanol as liquid 2 embodies a system with $v_2\ll v_1$ ($v_2 \approx 0.71 v_1$). 

\item[$v_2>v_1$:]  Taking pure water as liquid 1 and the ethanol-water mixture as liquid 2 creates a system with $v_2>v_1$ ($v_2 \approx 1.04 v_1$).  
\end{description}

Note that, for the first two cases,  the speed of sound of the ambient liquid is matched to that assumed by the scanner, as per the assumptions of our ray model.  This is not true for the third case, and so the ray model must be extended to account for the distortion of the near-face of the object; this is outlined in \ref{app:threemap}.

\section{Results: Circular Objects}  
\label{sec:circ}

We begin by using the ray model to map out the area distortion of circular objects, first for a simplest case of a linear array and then for our primary case of a curved array.  The latter will be validated against experimental images.  Unless otherwise stated, we assume the nominal parameters $r=4$ and $d=2$.  Note that length is subsequently presented in arbitrary units. 

\subsection{Linear Array}
\begin{figure}[t]
\centering
\textsf{(a)~~~~~~~~~~~~~~~~~~~~~~~~~~~~~~~~(b)~~~~~~~~~~~~~~~~~~~~~~~~~~~~~~~(c)}~~~~~~~~~~~~~~~~~~~~~~~~~~~~~~~~~~~~~~~ \\
\includegraphics[width=0.31\columnwidth]{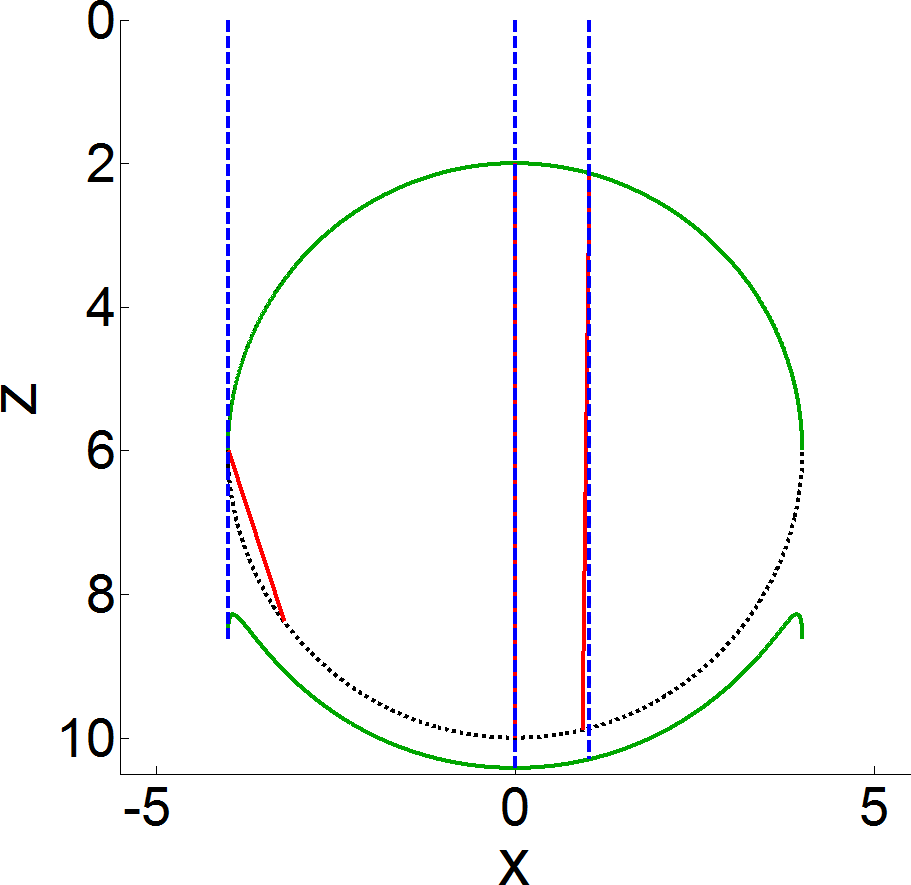}
\includegraphics[width=0.31\columnwidth]{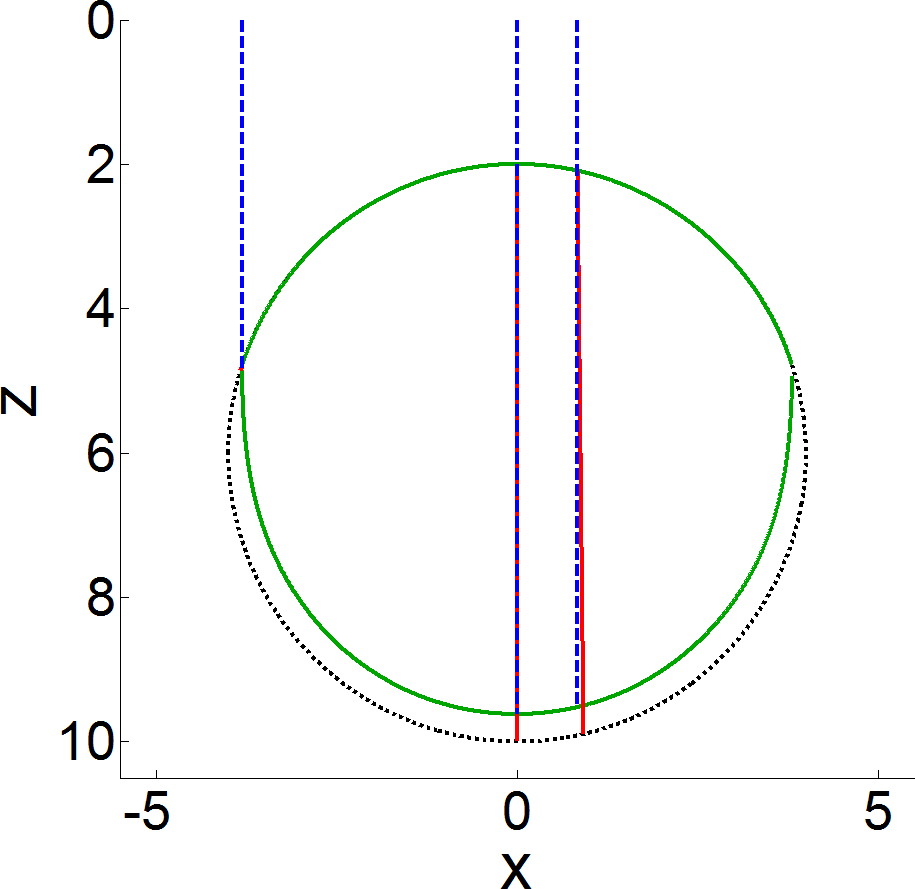}
\includegraphics[width=0.36\columnwidth]{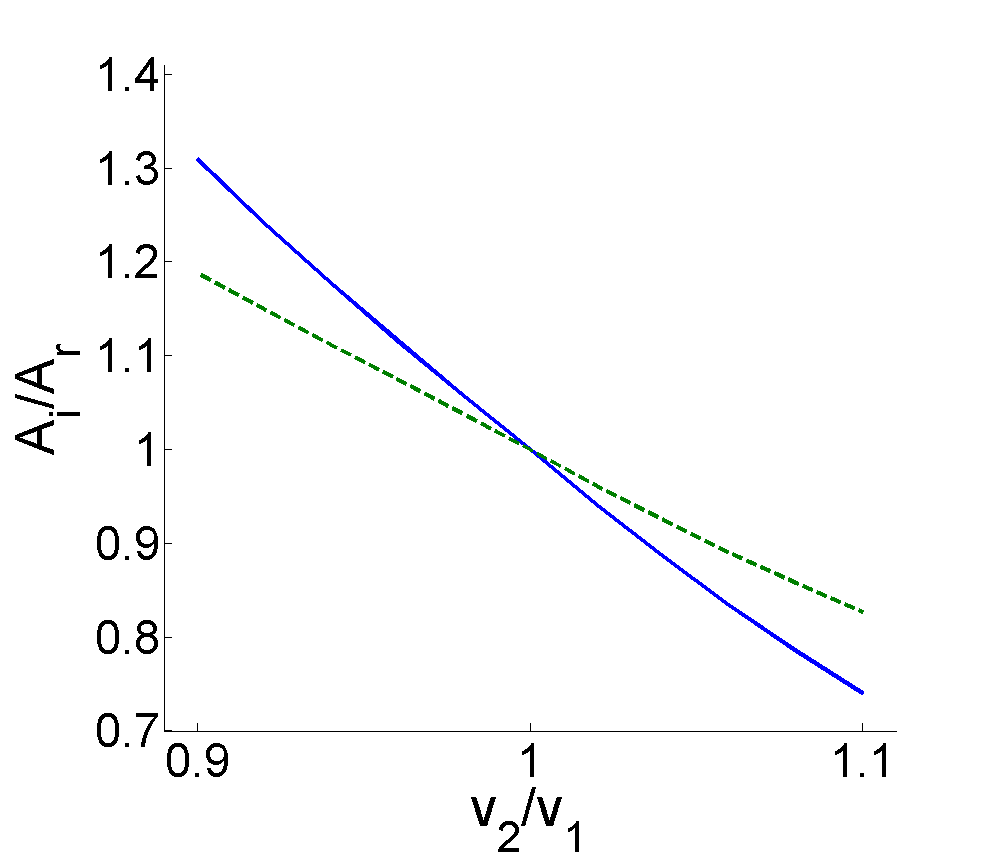}
\caption{(a)  Image (solid green line) of a circular object (dotted black line) with $v_2=0.95 v_1$ based on a linear array.  Three example sets of real rays (dashed blue lines) and their corresponding image rays (solid red lines) are overlaid.  (b) Same as above but for $v_2=1.05 v_1$.   (c)  Area distortion $A_{\rm i}/A_{\rm r}$ as a function of the speed of sound ratio  $v_2/v_1$, for the linear array (green dashed line) and curved array (blue solid line).}
\label{fig:linearcircle}
\end{figure}

In figure  \ref{fig:linearcircle}(a) we show the ray model prediction for $v_2/v_1=0.95$; the typical ray behaviour is illustrated by three sets of rays.  The rays propagate vertically downwards from the transducer surface (the $x$-axis, $z=0$), impinging the near-face (the upper half) of the object boundary.   Our assumption that the scanner is matched to the ambient speed of sound means that the near face is undistorted, that is, the image boundary (solid green line) and the object boundary (dotted black line) overlap here.  

At the near-face the real rays (red lines) are refracted inwards; the exception is the central ray which continues vertically due to its zero angle of incidence, $\theta_{\rm i}$.  These rays travel slower within the object since $v_2<v_1$.  The image rays (dashed blue lines) deviate in two ways: they are unrefracted at the near-face, and they appear elongated within the object.  The latter effect is best resolved along $x=0$: here the image ray is simply extended by a distance $2r(v_1/v_2-1)\approx0.42$.  Off-axis, the rays are refracted inwards (towards the normal), with the refraction being small across most of the object since $\theta_{\rm i}$ is small (with the deviation being hard to see by eye).  Towards the equator of the object, however, $\theta_{\rm i}$ grows rapidly, causing a significant increase in refraction and the real-ray path length in the object.  This has two notable effects as $x \rightarrow \pm r$: firstly, the increasing inwards refraction  leads to the appearance of a gap in the image boundary, and secondly, the divergence of the internal path length leads to the formation of prominent ``flicks'' in the image boundary \cite{Robinson}.

Now consider the case $v_2/v_1=1.05$, shown in  figure \ref{fig:linearcircle}(b).  Since $v_2>v_1$ the real rays travel faster than the image rays within the object, and so the far image boundary becomes shifted towards the transducer.  Also, the real rays become refracted outwards (away from the normal).  This removes the appearance of the above flicks for $x \approx \pm r$.  The image boundary again features a gap, although its origin is different to that above: here it is caused when the rays impinge the near face at a sufficiently large angle of incidence that they undergo total internal reflection, and so do not reach the far face.  

For the above cases the area distortion is $A_{\rm i}/A_{\rm r}=1.093$ and $0.907$, respectively.  In other words, a $5\%$ speed of sound difference has led to $\approx10\%$ error in the area.  We have also explored the wider dependencies of the area distortion on the key parameters.  The area distortion scales approximately linearly with $v_2/v_1$, as shown in figure \ref{fig:linearcircle}(c).  As $d$, the transducer-object distance, is varied, the image area $A_{\rm i}$ is unaffected, since all rays approach the object in the same direction.  Moreover, the image area scales in proportion to the size of the circular object, such that the area distortion ratio $A_{\rm i}/A_{\rm r}$ is independent of $r$, the radius of the object.

\subsection{Curved Array}
\label{sec:curved}
\begin{figure}[t]
\centering
\textsf{(a)~~~~~~~~~~~~~~~~~~~~~~~~~~~~~~~~(b)~~~~~~~~~~~~~~~~~~~~~~~~~~~~~~~(c)}~~~~~~~~~~~~~~~~~~~~~~~~~~~~~~~~~~~~~~~ \\
\includegraphics[width=0.30\columnwidth]{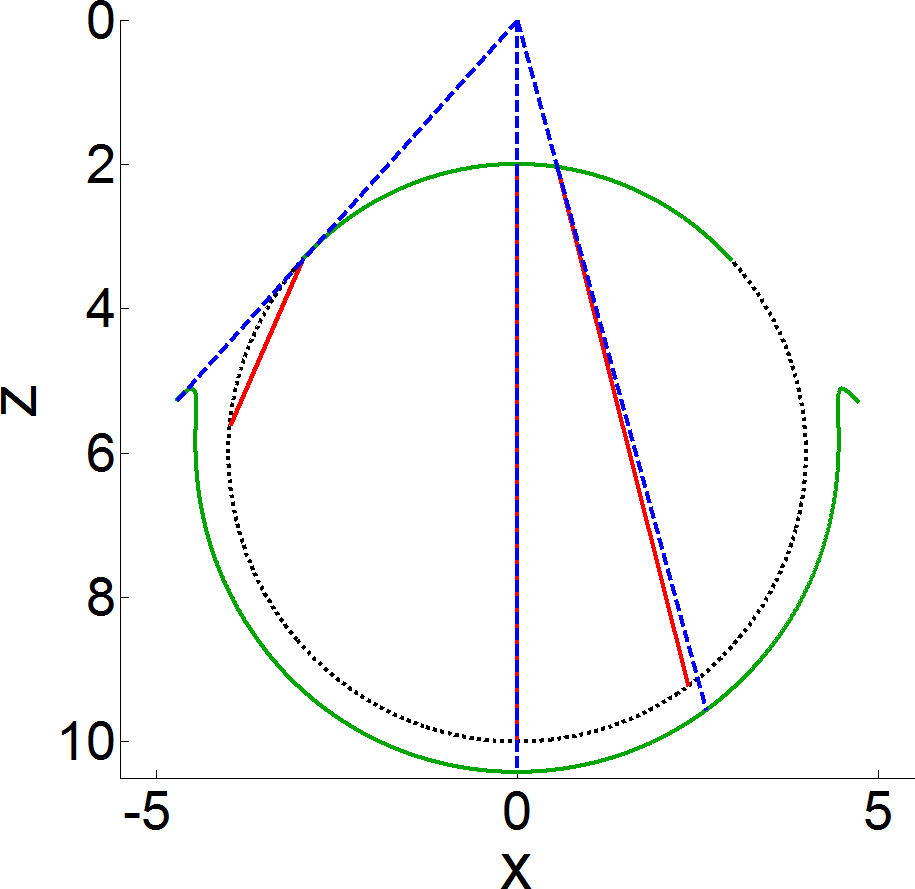}
\includegraphics[width=0.30\columnwidth]{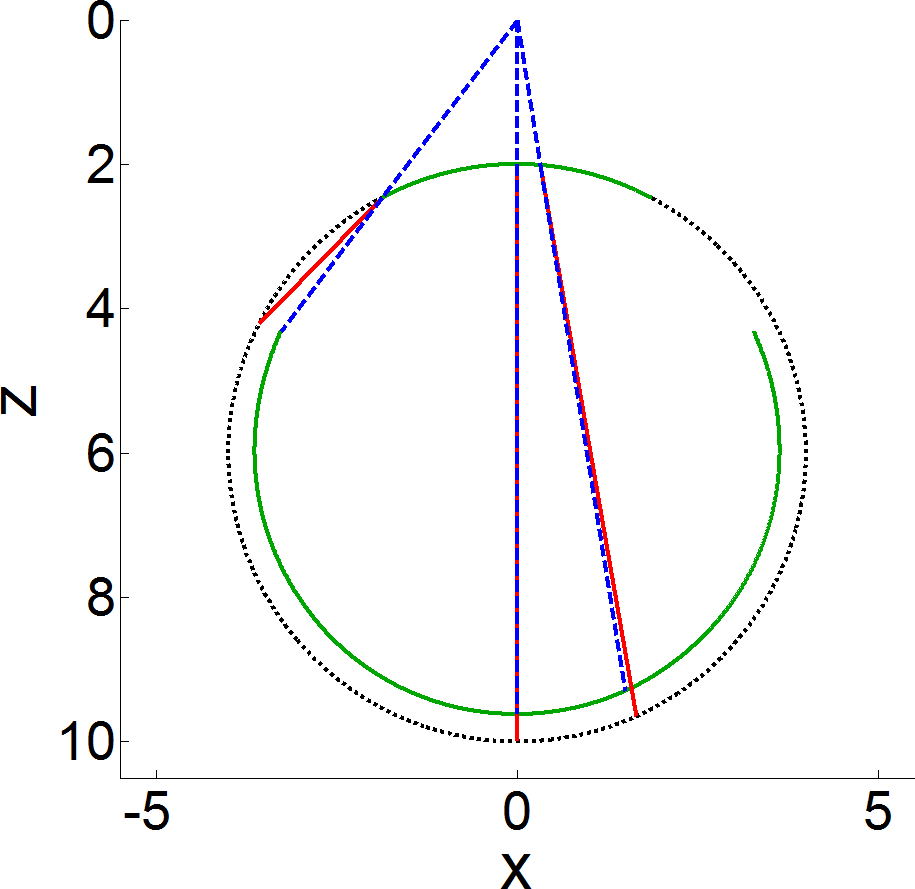}
\includegraphics[width=0.34\columnwidth]{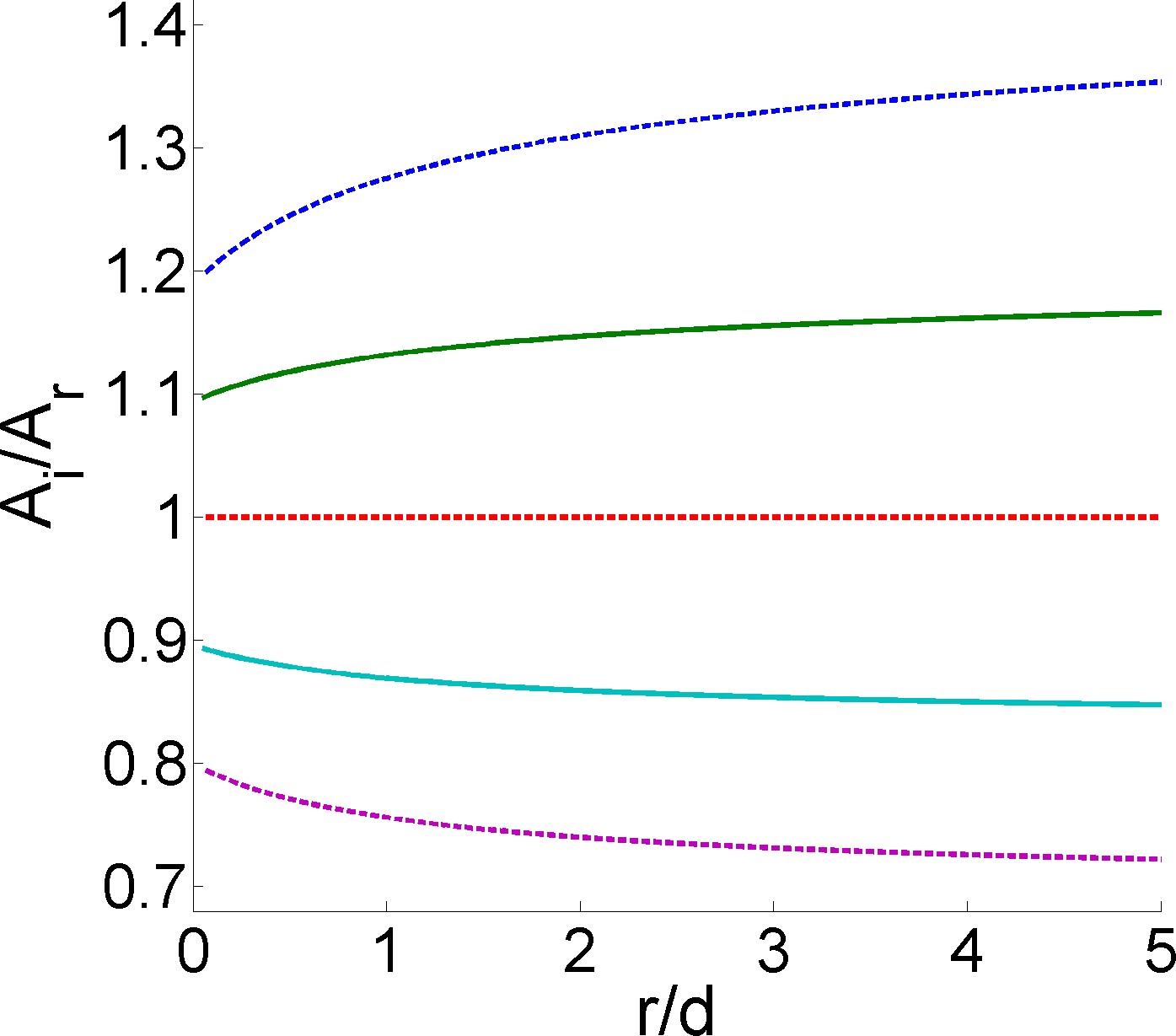}
\caption{(a)-(b) As figure \ref{fig:linearcircle} for a curved array (with $d+d_0=2$).  (c) Area distortion $A_{\rm i}/A_{\rm r}$ of the circular object under curved array imaging as a function of the ratio $r/d$.  Various values of $v_2/v_1$ are considered ranging from 0.9 (top line) to 1.1 (lowest line) in steps of 0.05.  The values $v_2/v_1=0.95$ and $1.05$ are highlighted by solid lines.}
\label{fig:curvedcircle}
\end{figure}

The corresponding images for a curved array are shown in figure \ref{fig:curvedcircle}(a-b).  We see similar qualitative effects as for the linear array: the shifting of the image boundary in front/behind the object boundary, gaps in the image boundary, and the appearance of flicks for $v_2 < v_1$.  However, the ray structure and image shape are modified further due to the angular spread of the emitted rays.  The near-face of the object has reduced extent, such that a greater proportion of the boundary is distorted.  The incident rays now approach the tangent of the object fore of the equator, which causes a corresponding shift in the gaps and flicks in the boundary towards the transducer.  For the $v_2<v_1$ case we find $A_{\rm i}/A_{\rm r}=1.147$; for $v_2>v_1$ we find $A_{\rm i}/A_{\rm r}=0.859$.  In other words, a $5\%$ difference in sound speeds for the curved array leads to over $14\%$ deviation in area.  

As for the linear array, the variation of the area distortion with $v_2/v_1$ is approximately linear [figure~\ref{fig:linearcircle}(c)], but with greater deviation than for the linear array.  The dependence on $r$ and $d$ is more complicated for the curved array since these parameters control the angular spread of the incident rays.  Figure \ref{fig:curvedcircle}(c) shows that the change in area distortion (relative to the undistorted value $A_{\rm i}/A_{\rm r}=1$) increases with $r/d$; this is due to the increasing angular spread of rays as $r/d$ increases, leading to greater refractive effects and increasing the proportion of the boundary which is distorted.

\subsection{Experimental Validation}

\begin{figure}
\centering
\textsf{(a)~~~~~~~~~~~~~~~~~~~~~~~~~~~~~~~~~~(b)~~~~~~~~~~~~~~~~~~~~~~~~~~~~~~~~~~(c)}~~~~~~~~~~~~~~~~~~~~~~~~~~~~~~~~~~~~~ \\
\includegraphics[width=0.31\columnwidth]{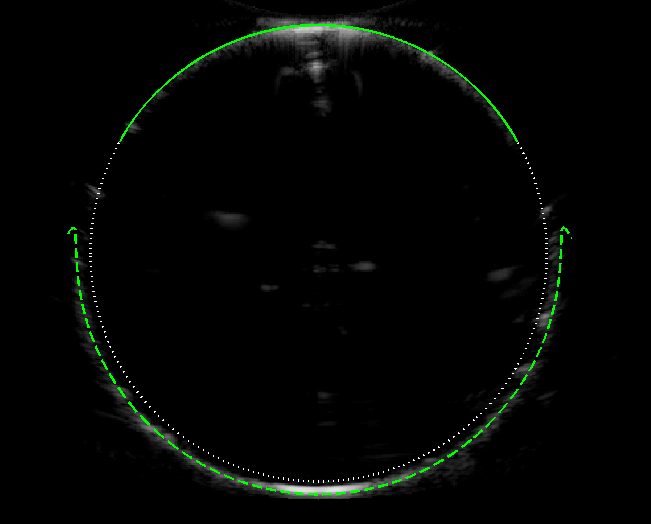}
\includegraphics[width=0.322\columnwidth]{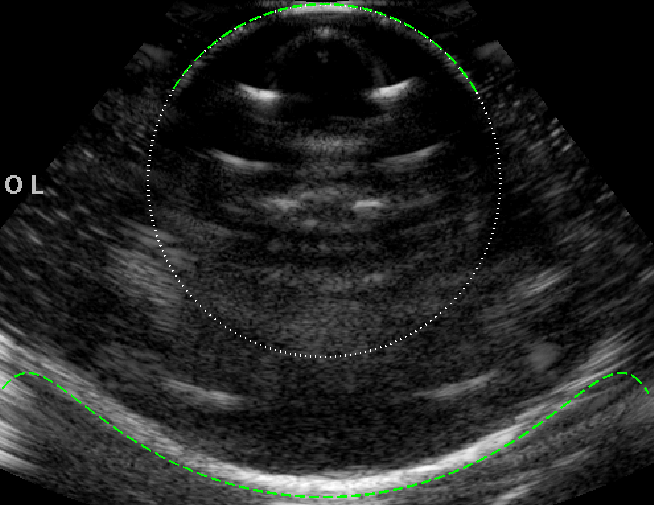}
\includegraphics[width=0.333\columnwidth]{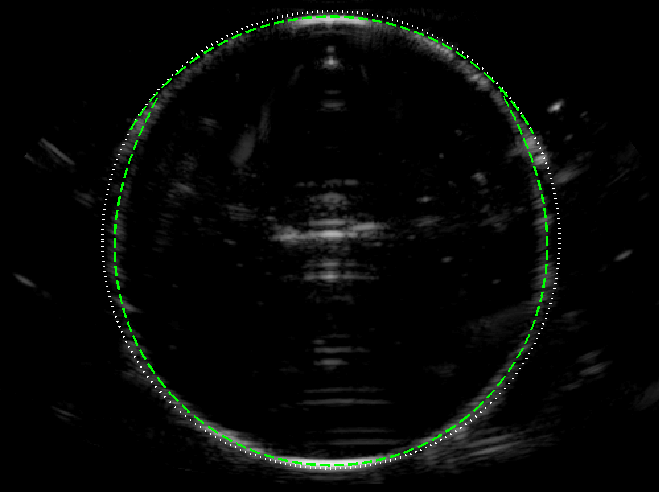}
\caption{B-mode images of the circular object of speed $v_2$ (imaged from above in a liquid with speed $v_1$), with the ray model superimposed (green dashed line). (a)  $v_2=1490$ m s$^{-1}$ and $v_1=1540$ m s$^{-1}$, giving $v_2/v_1=0.968$.  (b) $v_2=1100$ m s$^{-1}$ and $v_1=1540$ m s$^{-1}$, giving $v_2/v_1=0.71$.  (c) $v_1=1490$ m s$^{-1}$ and $v_2=1540$ m s$^{-1}$, giving $v_2/v_1=1.034$. 
}
\label{fig:exp3}
\end{figure}

Experimental B-mode images (with curved array) of the circular test object are shown in figure \ref{fig:exp3}, alongside the ray model predictions, for following 3 speed of sound scenarios.

\begin{description}
\item{(a) $v_2<v_1$:} Here we consider $v_1=1540$ m s$^{-1}$ (ethanol-water mixture) and $v_2=1490$ m s$^{-1}$ (water), giving $v_2/v_1=0.97$.  No geometric distortion is evident at the near face but the far face is shifted away from the transducer, in good agreement with the ray model (dashed green line).

\item{(b) $v_2 \ll v_1$:}  Now taking $v_2=1100$~m s$^{-1}$ (methanol) gives the more extreme case of $v_2/v_1=0.71$.  The far-face of the image is highly distorted, in good agreement with the ray model.

\item{(c) $v_2>v_1$:} Lastly, we take $v_1=1490$ m s$^{-1}$ (water) and $v_2=1540$ m s$^{-1}$ (ethanol-water mixture), giving $v_2/v_1=1.03$.  The ambient liquid is no longer matched to the scanner calibration, causing the near face to become slightly shifted {\it towards} the far face.  Again, the overall image shape agrees well with the ray model (in this case in based on the extended theory in Appendix A).
\end{description}
These results validate the use of the ray model in predicting the geometric distortion of the object boundary.

\section{Results: Elliptical Objects}
We extend our theoretical analysis to the more general case of elliptical objects, which allows us to explore the effect of elongation.  For brevity, we focus on the curved array.  

\begin{figure}[t]
\centering
\textsf{~~~(a)~~~~~~~~~~~~~~~~~~~~~~~~~~~~~~~~~~~~~(b)}~~~~~~~~~~~~~~~~~~~~~~~~~~~~~~~~~ \\
\includegraphics[width=0.25\columnwidth]{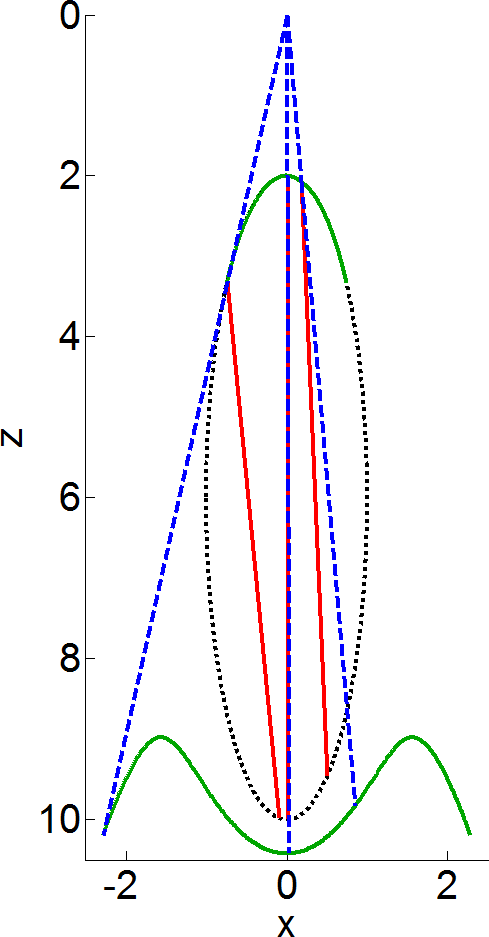}\hspace{1cm}
\includegraphics[width=0.25\columnwidth]{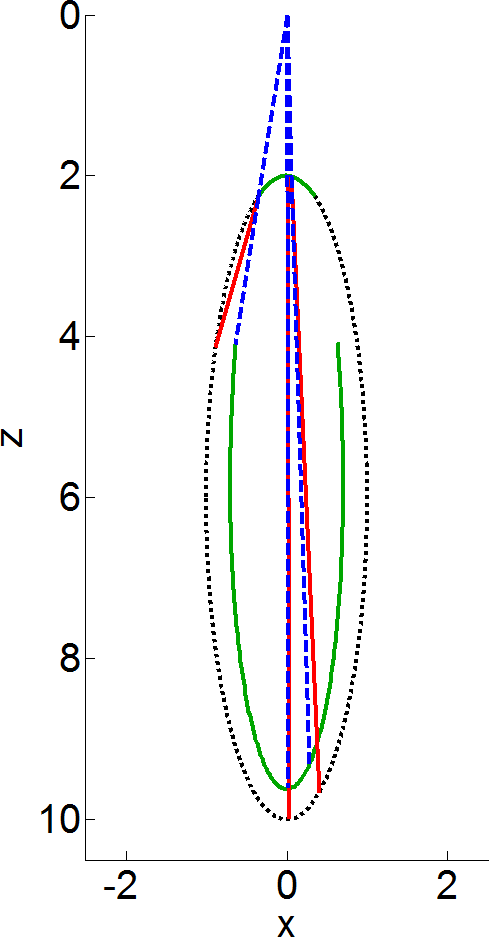}
\caption{Curved array images (solid green line) of an elongated elliptical object (dotted black line) with $b=1$ and $c=4$ ($d+d_0=2$).  Shown are the results for (a) $v_2=0.95 v_1$ and (b) $v_2=1.05v_1$.}
\label{fig:ellipse_elong}
\end{figure}
\begin{figure}[b]
\centering
\includegraphics[width=0.6\columnwidth]{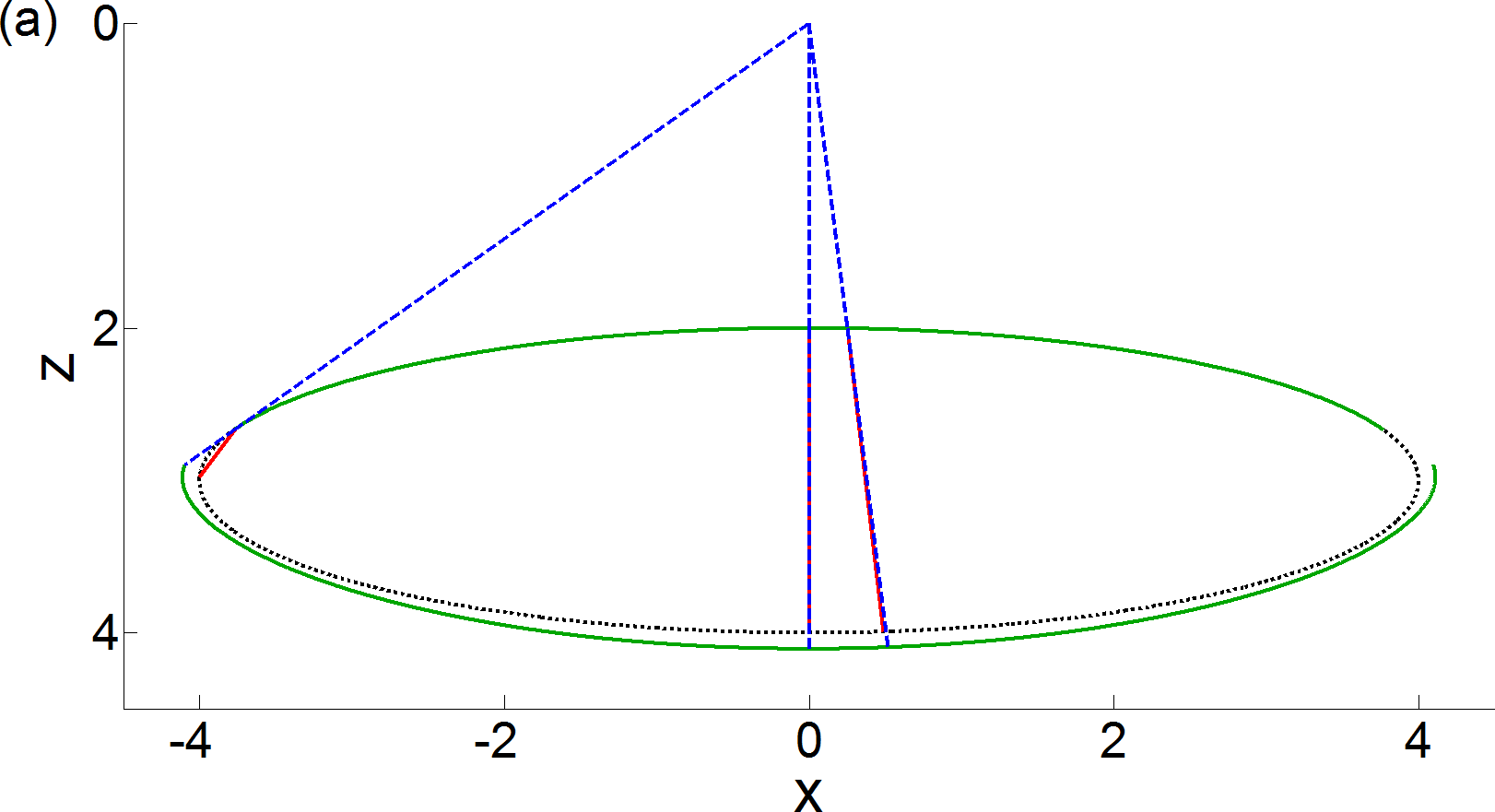}~~~~~~~~~~~~~~~~~~~~~~~~~~~~~~~~~~~~~~~~~~~~~
\\
\includegraphics[width=0.6\columnwidth]{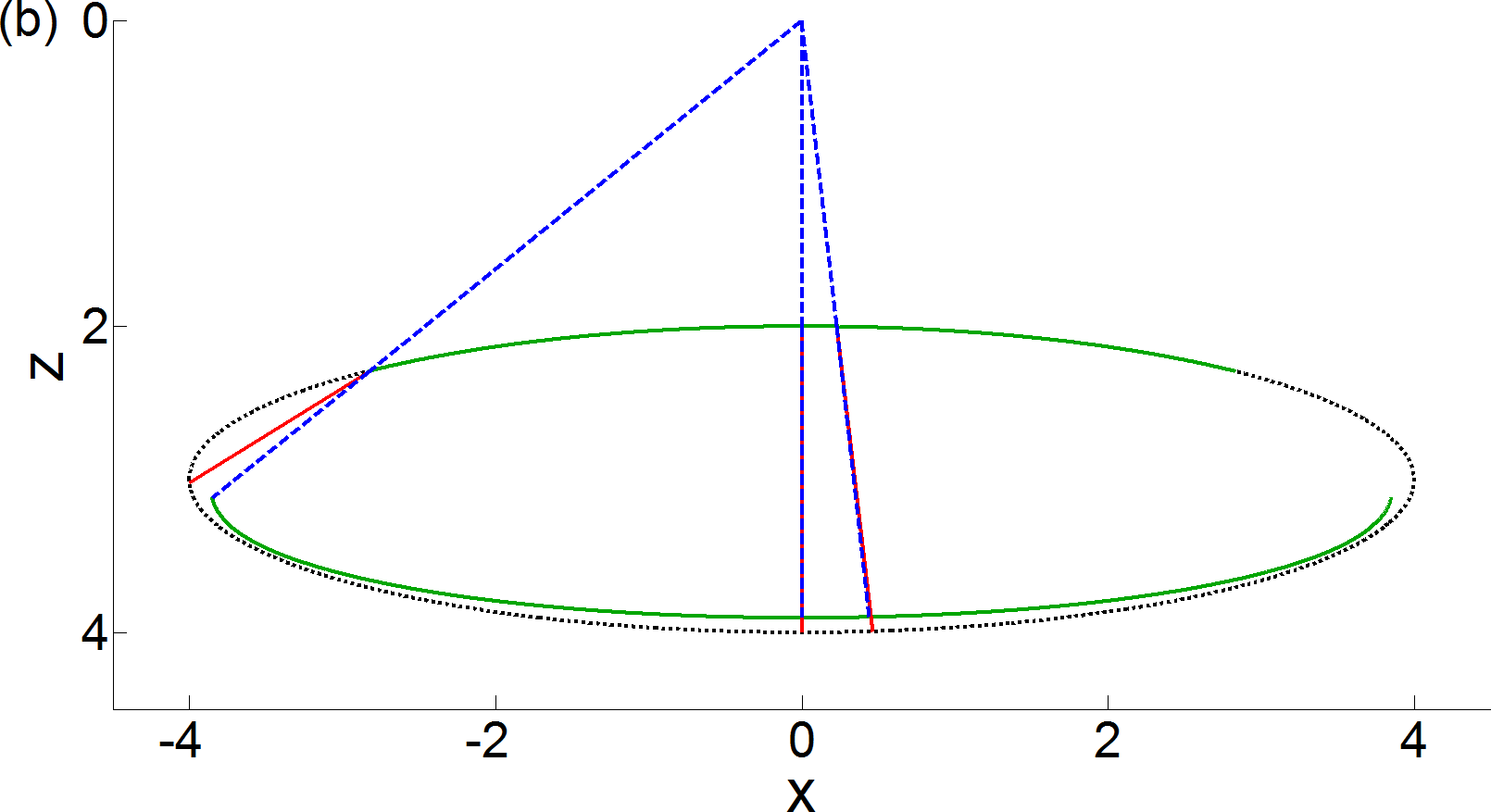}
\includegraphics[width=0.37\columnwidth]{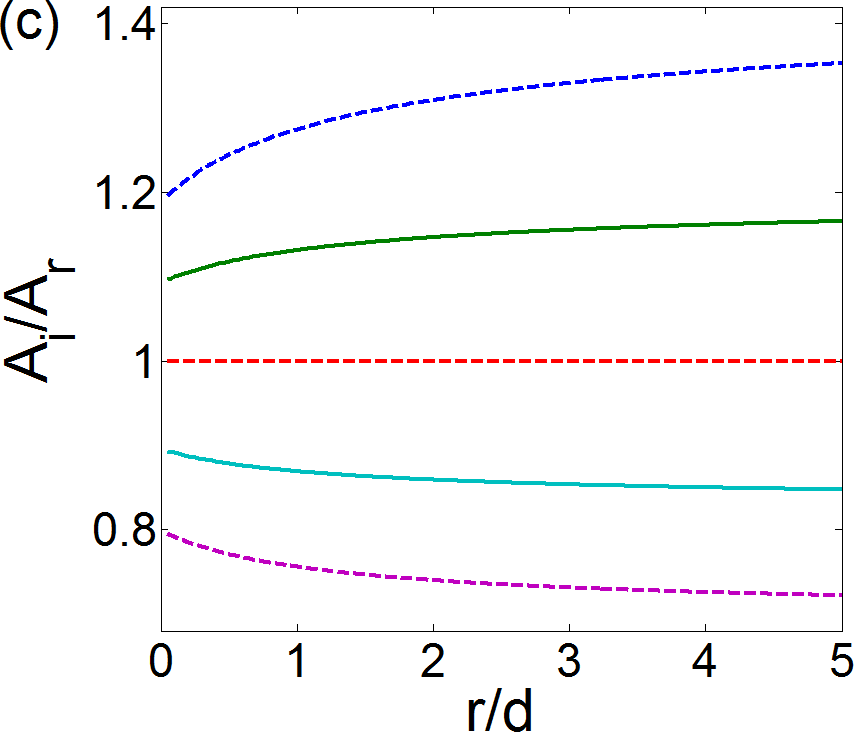}
\caption{(a)-(b) Same as figure \ref{fig:ellipse_elong} but for a flattened elliptical object with $b=4$ and $c=1$.  (c) Area distortion $A_{\rm i}/A_{\rm r}$ versus ellipticity $b/c$ (with $b$ fixed to unity) under curved array imaging.  Various values of $v_2/v_1$ are considered ranging from 0.9 (top line) to 1.1 (lowest line) in steps of 0.05.}
\label{fig:ellipse_flat}
\end{figure}

We first consider an elliptical object that is elongated along $z$, with nominal parameters $b=1$ and $c=4$.  The image distortion, presented in figure \ref{fig:ellipse_elong} for both (a) $v_2=0.95v_1$ and (b) $v_2=1.05v_1$, shows similar qualitative structure for the circular object [figure \ref{fig:curvedcircle}] but with enhanced distortion.  For $v_2<v_1$, the flicks are even more pronounced than for the circular case, caused by the fast roll-off of the object boundary with $x$ (which causes the angle of incidence to increase more quickly with $x$).   The area distortions are $A_{\rm i}/A_{\rm r} = 1.58$ (for $v_2/v_1<1$) and $0.68$ (for $v_2/v_1>1$); this is vastly greater than for circular objects in Section \ref{sec:curved}, demonstrating the sensitive role of elongation.

In contrast, for a flattened ellipse [figure \ref{fig:ellipse_flat} (a)-(b), nominal parameters $b=4$ and $c=1$] the image distortion is reduced, with negligible flicks appearing in the image due to the slower roll-off of the boundary with $x$.  Also, the narrowness of the object in $z$ means that the rays have little distance to accumulate any significant distortion.   The area distortions here are $A_{\rm i}/A_{\rm r} = 1.06$ (for $v_2/v_1<1$) and $0.95$ (for $v_2/v_1>1$), confirming the reduced distortion for flattened objects.

To further quantify the dependence on elongation, figure \ref{fig:ellipse_flat} (c) plots the area distortion $A_{\rm i}/A_{\rm r}$ as a function of ellipticity $b/c$.  We clearly see that the distortion diverges for objects which are increasingly elongated (along $z$), and converges towards the undistorted result for increasingly flattened geometries.

\section{Discussion and Conclusions}
\label{Discuss}

We have developed a ray model that describes the geometric distortion of two-dimensional ultrasound images (based on linear, curved, phased or vector arrays) of circular and elliptical objects due to speed of sound discrepancies.  The geometric distortion arises from the refraction of the sound and the misrepresentation of length within the object.  By comparing to ultrasound images of a test object we have validated that the ray model successfully captures the underlying geometric distortion of the image.  

The geometric distortion of ultrasound images of circular objects has been examined previously \cite{Ziskin1,Ziskin2}, including by ray models \cite{Robinson,Sommer}.  Our analysis extends this previous body of work by {\it quantifying} the distortion of area, and thereby providing new insight into the errors in the ultrasonic evaluation of area and volume of anatomical features. Our work also extends previous analyses by considering elliptical objects, which allows us to explore the role of elongation.  

When $v_2<v_1$, the image is larger than the object, and vice versa for $v_2>v_1$.  The ray model predicts the emergence of gaps in the image boundary, an effect which may contribute to the phenomenon of acoustic shadowing \cite{Robinson,Ziskin1,Ziskin2,Sommer}.  For $v_2<v_1$ the image also features distinctive flicks at the extreme angular positions, as noted previously \cite{Robinson}.  The deviation of the image area from the true value is greater for the ``angled" array systems (phased, curved and vector arrays) than the linear array, due to the enhanced role of refraction.  The deviation also depends on the shape and size of the object, in general increasing with the lateral size of the object and its elongation along the ultrasound axis.  

One might expect that the area distortion follows the fractional mismatch in the speed of sound \cite{Ziskin2}.  However, we typically find the distortion to be considerably larger.  Take, for example, the circular object case considered earlier: there a  5\% mismatch in the speeds of sound led to a 9\% error in the area for a linear array and over 14\% error for a curved array.    The area distortion has contributions from refraction of the rays at the near surface and the miscalculation of the length of the ray within the object.  We can artificially remove the effect of refraction from the ray model by setting $\theta_{\rm r}=\theta_{\rm i}$.  In doing so we find for a circular object that these effects roughly contribute to half of the area distortion (i.e., refraction accounts for 40-50\% of the distortion); more generally this proportion will depend on the object shape. 

As well as geometric distortions, ultrasound images undergo well-known intensity distortions, arising from wave behaviour and the true beam line structure \cite{Steel2004}.  In future it would be insightful to consolidate these two distortion contributions by direct comparison between ray predictions and more sophisticated finite-element models of the full acoustic field.

Our work has considered objects with known areas and derived their image.  Of more importance in the clinic is the inverse problem, that is, to obtain the true area given an image, and developments in this direction could promote algorithms to correct the distortion in clinical measurements.  A further extension of this work would be to investigate the volumetric distortion of 3D objects with atypical speed of sound.  Due to the increased dimensionality of this situation, one would expect the distortion to be further amplified compared to the 2D case.  With ultrasound-based volume measurements now claiming to have errors of a few percent \cite{Kristiansen,Treece_2001,Tong_2006}, these distortions may become the significant contributor to volumetric errors.

 \appendix

 \section{Ray Model for $v_1 \neq v_0$}
 \label{app:threemap}

For simplicity our main ray model considered the ambient speed of sound $v_1$ to be matched to that of the scanner $v_0$, thereby negating the misrepresentation of distance as the ray travels through the ambient liquid.  Here we extend the ray model to account for $v_1 \neq v_0$ with a curved array.

\begin{figure}[!htb]
\centering
\includegraphics[width=0.45\columnwidth]{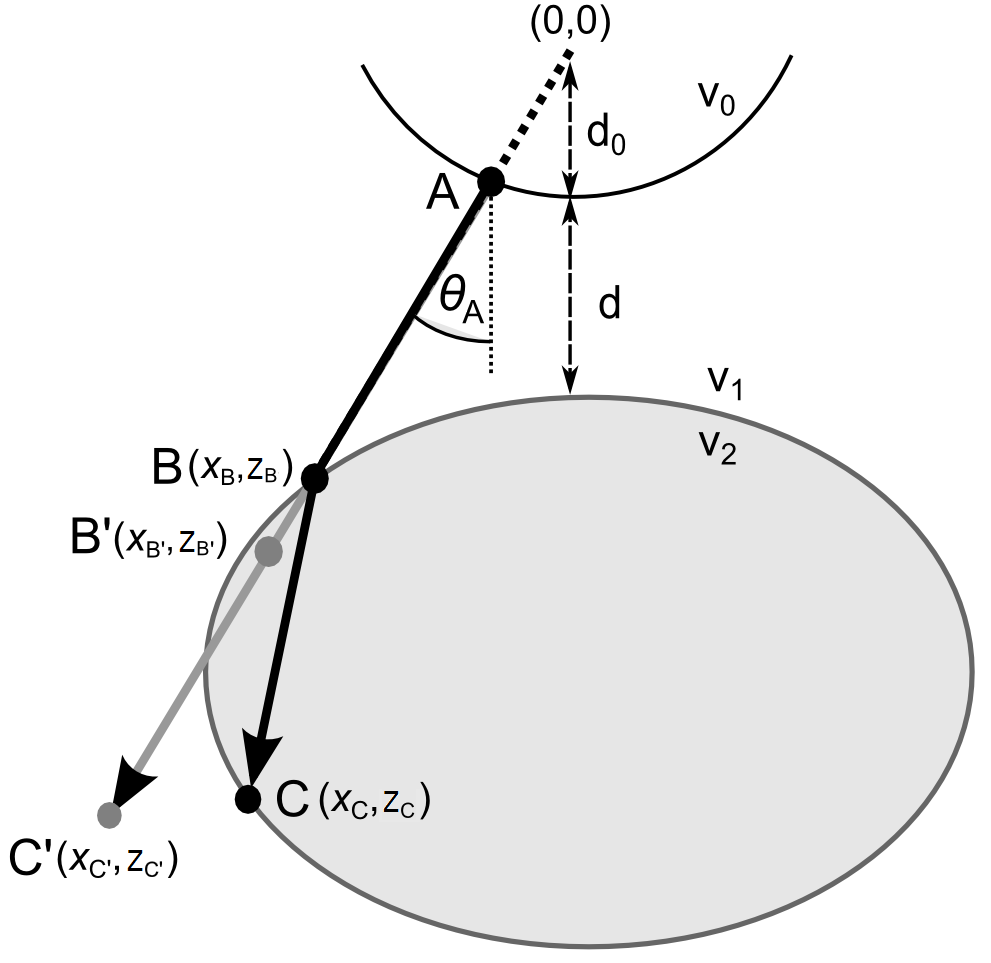}
\caption{Schematic of the transducer-object scenario, featuring an elliptical object (speed of sound $v_2$) within an ambient medium (speed of sound $v_1$).   The transducer is calibrated to a speed of sound $v_0$.}
\label{fig:threediagram}
\end{figure}

We consider a transducer element on the surface of the curved array, point A [figure~\ref{fig:threediagram}].  Since the ray is emitted {\it normal} to the surface, no refraction effects occur at the transducer boundary, despite $v_1 \neq v_0$.  The path of the real ray, which refracts into the object at point B on the near face and reaches the far boundary at point C, is provided by the same theory as before. The difference, however, now arises in the position of point B and its image point B'.  We add a misregistration to all of the points by introducing the two relations for the misplacement of $x$ and $z$, namely:
\begin{eqnarray*}
\fl \Delta X=\left(\frac{v_0}{v_1}-1\right)\left[\sqrt{x^2+z^2}-d_0\right]\sin\theta_{\rm A};~~
\Delta Z=\left(\frac{v_0}{v_1}-1\right)\left[\sqrt{x^2+z^2}-d_0\right]\cos\theta_{\rm A},
\label{eq:Dc}
\end{eqnarray*}
which are similar to those in Ref. \cite{Goldstein2000}.  Thus the position of the image point B' is
\begin{eqnarray*}
(x_{\rm B'},z_{\rm B'})&=&(x_{\rm B}+\Delta X,z_{\rm B}+\Delta Z).
\label{eq:circleimage2}
\end{eqnarray*}
The total length of the virtual ray, the equivalent of equation (\ref{eq:length}), is
\begin{eqnarray*}
L_{\rm AC'}=\frac{v_0}{v_1}L_{\rm AB}+\frac{v_0}{v_2} L_{\rm BC},
\label{eq:length2}
\end{eqnarray*}
and the image point C' is located at
\begin{eqnarray*}
(x_{\rm C'},z_{\rm C'})=(L_{\rm AC'} \sin\theta_{\rm A}, L_{\rm AC'} \cos\theta_{\rm A}).
\end{eqnarray*}
Thus the distortion of the near-face and the far-face is completely determined.

\end{document}